\documentstyle[twocolumn,epsfig,prl,aps,amsmath,amssymb]{revtex}

\begin{document}
\twocolumn[\hsize\textwidth\columnwidth\hsize\csname @twocolumnfalse\endcsname
%\tightenlines
%\begin{normalsize}

\title{Manifestation of marginal Fermi liquid and phonon excitations in 
photoemision experiments
 of cuprate superconductors}

\author{{A. Greco} and {A. Dobry}} 

\address{
Departamento de F\'{\i}sica, Facultad de
Ciencias Exactas e Ingenier\'{\i}a and \\
IFIR(UNR-CONICET), Av. Pellegrini 250, 
2000-Rosario, Argentina
}

\date{\today}
\maketitle
%\vspace{5cm}

\begin{abstract} 
Recent ARPES experiments in cuprates superconductors
 show a kink  in the electron dispersion  
near the Fermi energy.
This kink coexists with a linear frequency dependence
 of the imaginary part of the electron self-energy.
In this paper we show that both features could be accounted for
if an electron-phonon interaction is included in a model where the
electrons are described  
by a marginal Fermi liquid theory. Phonons provide the energy scale seen
in the experiments but the quasiparticle weight at the Fermi level
is zero. At high binding energy, in agreement with the experiment,
the electron dispersion
does not go to the one-electron band.
 We analyze the compatibility between the 
electron scattering rate extracted from ARPES experiment and the 
one extracted from transport properties. We conclude that the 
 electron-phonon interaction relevant for transport properties
is strongly screened respect to the one extracted from ARPES. 
This is in agreement with recent studies in the context of
$1/N$ expansion on t-J model.
\end{abstract}
 
%]
\vskip2pc]
%\newpage

Recently, the improvement of momentum resolution in ARPES
has generated new discussions about the interpretation of the electronic 
dynamic and its relation with the phenomena of high-$Tc$ 
superconductivity. 

In one of these experiments \cite{Valla} the energy distribution 
curves (EDC) and momentum distribution curves (MDC)
in Bi2212 (optimally doped) gave important information about the 
electron self-energy $\Sigma(\vec{k},\omega)$. 
The main result of \cite{Valla}  
was related with the suitable linear frequency dependence of the 
$-Im(\Sigma(\vec{k},\omega))$ with no saturation 
up to frequencies of the order of $150 meV$. 
Surprisingly, this result is very similar to the one already predicted
 in the early days of the HTC superconductivity \cite{Varma}.  
In particular, Anderson \cite{Anderson} 
and Abrahams and Varma \cite{Abrahams} related these new results 
with their original ideas of spin-charge separation 
and marginal Fermi liquid (MFL)
theory respectively. 
 
However, the subsequent ARPES experiments in Bi2212 \cite{Bogdanov,Kaminski} 
showed a new picture. Both papers found the presence 
of a kink in the electron dispersion near the  Fermi surface (FS). 
This kink is basically a change in  the slope of the dispersion, near the 
FS, by a factor of two and is related with an energy scale of the order 
of $50 meV$.
It is interesting to note that this low energy scale was not discussed 
in \cite{Valla} but a close inspection of this paper 
shows that the kink was also observed there
(see Fig.1 of \cite{Valla})

A straightforward interpretation for the 
break in the electron dispersion is associated with a change in the Fermi 
velocities (FV), near the FS, by a factor of two. This can 
be related with a quasiparticle weight $z=1/(1+\lambda) \sim 0.5$,
 giving a characteristic coupling $\lambda \sim 1$.
But if $z$ is 
finite, the system behaves like a Fermi liquid (FL).
On the other hand, as we mentioned above, the experiment
shows that $-Im(\Sigma(\vec{k},\omega))$ is linear in
$\omega$ implying a zero quasiparticle weight at the FS 
which is not consistent with a FL picture.
Therefore, it is important to investigate whether the kink 
is consistent only with a FL picture or it can also be 
interpreted in terms of  non Fermi liquid theories.

The presence of this kink, which nowadays is well accepted, opens several and
relevant questions: which mechanism provides the energy scale of the kink?;
is $-Im(\Sigma(\vec{k},\omega))$ really  linear in $\omega$
 for all $\omega$ or, it deviates near the FS?;
if so, is spin-charge separation or MFL the right scenario
to account for this low energy breakdown of the linear behavior?.

Further experiments \cite{Lanzara} showed the kink also in monolayer
compounds LSCO and Bi2201 for several dopings (underdoped, optimally doped and 
overdoped). It was also shown in \cite{Bogdanov,Johnson} that 
the kink and the self energy effects are rather isotropic through 
the Brillouin zone (BZ) and it exists below and above  $T_c$.
Another interesting feature of the kink is its evolution with  
temperature $T$. When $T$ increases the change of the slope of the dispersion 
decreases, making the kink less evident.

There are two main explanations for the kink. The first one 
 relates the kink 
with a magnetic excitation\cite{Eschrig} which appears as a resonant peak 
in neutron scattering measurements 
\cite{Fong} and the second one associates the kink with the electron-phonon
interaction(EPHI) \cite{Zeyher,Shen}. 
In Ref. \cite{Zeyher} using a large-N expansion on the t-J model,
it was shown that phonons are better candidates than the resonant 
peak to explain the low energy features associated with the kink.
Besides, from the experimental point of view, 
as it was mentioned above,
the kink is observed in optimally and overdoped regime above $Tc$ 
and also in monolayer compounds. 
As the resonant peak was only observed in the superconducting phase of
the  bilayer compounds, it is problematic to associate the kink with
this magnetic resonance. 
Therefore, in the present paper we will take the EPHI
\cite{Zeyher,Shen} as the origin of the kink. 
   
At this point we note that an EPHI can not explain the linear 
behavior of $-Im(\Sigma)$, measured in ARPES, with no saturation 
up to frequencies much larger than the Debye frequency. 
Recently this linear behavior 
in $-Im(\Sigma)$ was experimentally 
re-investigated for several dopings 
from under to overdoped \cite{Yusof}.
 
One important characteristic of the electron dispersion   
 is that, in contrast to usual metals \cite{Mahan,Valla1},
 does not go to the one-electron band predicted from LDA 
calculation up to the measured energy of $200 meV$ (much larger than the 
Debye energy).
This was remarked only in
\cite{Bogdanov} (to our knowledge).  
It will be a very important point in the present paper.
Moreover, we consider that this is a clear manifestation of the existence 
of large $-Im(\Sigma)$ at large $\omega$, as observed in Bi2212.

In this paper we will analyze the latest ARPES experiments in the context 
of the MFL. We will show that the MFL with an additional electron-phonon 
interaction can reproduce the ARPES results quite well.   
Our calculation will show that a slope-change
in the dispersion of order two 
is also compatible with a theory that predicts a zero quasiparticle weight $z$ 
at the FE. We point out that our analysis will be based manly on 
phenomenological grounds. It means that we will not solve or treat 
rigorously any microscopic Hamiltonian.

We will consider the following total self-energy
$\Sigma(\vec{k},\omega)=\Sigma_{MFL}(\vec{k},\omega)+\Sigma_{e-ph}(
\vec{k},\omega)$
with $\Sigma_{MFL}(\vec{k},\omega)$  the MFL self energy 
\cite{Varma} given by
$\Sigma_{MFL}(\vec{k},\omega)=\lambda[\omega log(\frac{x}{\omega_c}-i 
\frac{\pi}{2} x)]$.
$x=max(|\omega|,T)$ and $\lambda$ is a coupling constant (as we are
interested in the general behavior we neglect, as a first approximation,
possible vertex corrections). 
An important characteristic of the MFL theory is the isotropic 
nature of the self-energy.
$\Sigma_{e-ph}$ is the contribution to the self energy by the 
EPHI. 
The result for $\Sigma_{e-ph}$ is not sensitive weather we use
in its calculation the MFL Green function 
($G_{MFL}=1/(i\omega_n-(\epsilon(k)-\Sigma_{MFL})$)  
or the simple bare electron Green function. 
This result is more evident 
when the cutoff $\omega_c$ increases.  
According to this remark, in the following we will present results for 
$\Sigma_{e-ph}$ using the bare Green function. 

In the following we will consider a Holstein model,
with Debye frequency $\omega_{D}$, for 
phonons and a $\vec{k}$-independent electron-phonon coupling $\lambda^{ph}$.
We will also calculate $\Sigma_{e-ph}$ assuming a constant 
electronic density of state (DOS).
A more accurate calculation, which include a more realistic model 
for the electrons and phonons, will not change the main conclusion 
of the present paper. 

We also need the one-electron band $\epsilon(\vec{k})$. 
For $\epsilon(\vec{k})$ we assume 
$\epsilon(\vec{k})=-2t(cos(k_x)+cos(k_y))+4t'cos(k_x)cos(k_y)$ with $t=150meV$ 
and 
$t'/t=0.25$. With these parameters $\epsilon(\vec{k})$
 reproduces the FS of Bi2212 \cite{Norman}.
The small value of these hopping parameters means that they are
renormalized by electronic correlations \cite{Moreo,Zeyher}.
We will present calculations for doping $\delta=0.20$ away from half-filling.

In Fig.1 we show the imaginary part of the total electron 
self-energy for the parameters
$\lambda^{ph}=0.75$, $\omega_D=34.5 meV$, $\omega_c=1.8 eV$ and
$\lambda=(2/\pi)0.55$. We choose the parameters in order to reproduce 
the experimental results of \cite{Valla}. 
Moreover, we fix the parameters
in order to reproduce the difference, at zero frequency 
between $-Im(\Sigma)$ at $300K$ and $90K$. Using this criteria 
we avoid the uncertainty due to the contribution of 
the impurities in the samples. 

The agreement between the results of Fig. 1 
and the experiments is quite good with characteristic phonon 
parameters consistent with those
previously  proposed\cite{Zeyher,Andersen}.  
On the other hand we can see in Fig. 1 the overlap between the results 
for $T=300K$ and $90K$ for frequencies larger than $60 meV$. 
This fact was experimentally observed and 
 put strong constraints on the values of the $\lambda^{ph}$ and $\omega_D$. 
Smaller values of these two parameters will not reproduce 
the mentioned overlap. 
In the inset of the figure we 
show the behavior of $-Im(\Sigma_{MFL})$ for
a pure marginal Fermi liquid model at the same two temperatures. 
We can see 
the qualitatively similar behavior in $\omega$ between the inset and the main 
figure. Therefore as was discussed in Ref. \cite{Valla},
 it is possible to describe the experimental result
with a pure MFL model by increasing   
 $\lambda$ from  
our value of $\lambda=(2/\pi)0.55$ \cite{Valla}. 
We conclude that  is not possible
 to distinguish between the two models by solely fitting
$-Im(\Sigma)$.
We need, in addition the $Re(\Sigma)$. This information enters 
directly in the kink or break of the dispersion near the FS as we will show
later.
Nevertheless, if it were possible to measure $-Im(\Sigma)$ for larger
energies than the ones currently available, our model predicts
a splitting between the results for $-Im(\Sigma)$
corresponding to this two temperatures for $\omega >> \omega_D$. 
This should be a possible experimental check of the present model.
 
In Fig.2 we present results for the electron dispersion near the FS in
the nodal direction.
This dispersion was obtained as the energy position of the main peak,
for a given $\vec{k}$, of the spectral function calculated with the 
self energy correction previously discussed. 
In Fig. 2 the kink is clearly observed (diamonds) at 50K
near the FS. For large binding energy the electron dispersion
(diamonds) does not go to the one-electron band (circles).
It is also clear that, when the temperature 
increases from 50K to 130K, the kink is less evident
 like in the experiment.
At larger energies than the Debye frequencies, the band goes to 
the pure MFL results. 
The figure also shows, at small temperatures, a change in the slope 
of the dispersion consistent with the experiment. 
This also means that our phonon parameters have the correct order of 
magnitude.
The inset of the figure
 shows the electron dispersion corresponding to  MFL at two 
different temperatures.
From this inset we conclude that the  pure MFL theory cannot   
reproduce the kink. There is only a small 
deviation (if the temperature is large) near the FS but in the 
opposite direction to the experiment.

From the results shown in Fig. 2 we conclude that: to explain the kink 
in the context of MFL we really need an EPHI which provides 
an energy scale of the order of $50 meV$. 
The presence of an EPHI to explain the kink does not mean 
a weakness of the MFL, on the contrary, we think that the MFL 
is an  excellent phenomenological insight to 
discuss the physics of high-$T_c$ cuprates.

In Fig. 3a we plot the inverse of the quasiparticle weight
$Z=1-(Re(\Sigma(\omega))-Re(\Sigma(0)))/\omega$ 
($Z=z^{-1}$) for  the 
case of only MFL $Z_{MFL}$ (dashed line), only phonons $Z_{ph}$ 
(dot-dashed line) and the total case $Z$ (solid line). 
As it is well know,  the quasiparticle weight goes to zero logarithmically 
at zero $\omega$ for the pure MFL. 
The solid line shows that $Z$ goes to infinity (the quasiparticle weight 
goes to zero) logarithmically but also shows the structure 
at $\omega=\omega_D$ which is the responsible for the kink.
From Fig. 3a, we can say that a change in the 
slope by a factor of two does not means 
that we really have a FL picture. 
In our case we can reproduce the main features of ARPES 
together with a zero quasiparticle weight at the FS. 

One relevant question for high-$Tc$ cuprates is weather the relaxation time 
$1/\tau$ measured in ARPES determines the transport properties. 
The $DC$ resistivity can be calculated as 
$\rho=4\pi / {\Omega_p}^2 \tau_{tr}$ where, in usual metals
at high temperature,
$1/\tau_{tr}= 2 \pi \lambda_{tr}^{ph} T$ \cite{Mahan,Burgy}. On the other 
hand, in most metals \cite{Allen} the electron-phonon coupling 
constant $\lambda^{ph}$ is nearly the same that $\lambda_{tr}^{ph}$ and then, 
the relaxation time measured in ARPES is the same that enters in transport.
The width of the Drude peak in the optical conductivity determines 
the quasi-particle lifetime called $1/\tau^*$. For the case of EPHI in
usual metals $1/\tau_{ph}^*=2\pi \lambda_{tr}^{ph}/(1+\lambda^{ph}) T$. 
The experiment in cuprates\cite{Schlesinger} shows that 
$1/\tau^*=1.5 T$. Our previous results show that an electron-phonon coupling 
$\lambda^{ph}$ of order $0.75$ is needed in order to reproduce ARPES.
If we consider, as in usual metals, 
that  for HTC cuprates $\lambda^{ph}=\lambda_{tr}^{ph}$ , 
we found for $1/\tau^* = 2.7 T$. Then, taking into account 
only the electron-phonon contribution we found  twice the slope 
of the experimental result. We have in addition a contribution 
to $1/ \tau^*$ from the MFL which, will increase even more  
the discrepancies with the experiment.

Then, to explain optical data we need $\lambda_{tr}^{ph} << \lambda^{ph}$.
Phonons are expected to be isotropic on the BZ 
and this is the reason because in usual metals
 $\lambda^{ph}=\lambda_{tr}^{ph}$. 
In their original paper Zeyher and Kuli$\acute{c}$ \cite{Kulic}
have shown $\lambda_{tr}^{ph} << \lambda^{ph}$ 
when the electrons are strongly correlated. 
The electron correlations are the responsible for vertex corrections of 
the EPHI. 
This vertex correction produces a $\vec{k}$-dependence in the final 
electron-phonon coupling (which was originally isotropic).
As a final result, vertex corrections favor 
forward scattering which leads to a $\lambda_{tr}^{ph}$ much smaller 
than $\lambda^{ph}$. 
In Ref. \cite{Greco} it was shown that 
$\lambda_{tr}^{ph}$ 
renormalizes to around $\lambda/3$.
Using a value of $\lambda_{tr}^{ph} =0.3$ and $\lambda^{ph}=0.75$ we obtain
$1/{\tau^*}_{ph}=1.07 T$. We have an additional contribution from 
the MFL, which for our set of parameters gives $1/{\tau^*}_{MFL}=1.09 T$. 
Then, the addition of both 
contributions ($1/{\tau^*}_{ph} +1/{\tau^*}_{MFL}$) 
give a value in good agreement with the experiments.

There are
two contributions to the resistivity, 
$\rho=(8\pi/{\Omega_P}^2) (\lambda^{ph}_{tr} +\lambda^{MFL}_{tr}) T$. 
According to our previous discussions we will take for the transport
electron-phonon coupling 
$\lambda^{ph}_{tr}=0.3$. Assuming the isotropic nature of the MFL we have
$\lambda^{MFL}_{tr}=0.55$.
With these two parameters and the experimental value $\Omega_p=3 eV$
we show (with circles) the temperature dependence of the DC resistivity 
in Fig 3b.

Our calculation compares well with 
 overdoped samples (see figure 4 of\cite{Yusof} for Bi2212).
It is evident from this experimental data that 
the resistivity is not linear in $T$ at less for low temperatures, near $Tc$.
It has a weak low temperature curvature similar to our results (circles)
 of Fig. 3b.
The weakly changes between the slopes at
 low $T$ given by the MFL 
and the slope at large $T$,  
is an additional manifestation of the small electron-phonon 
contribution to the transport properties.

It is very important to note that the same sample, used in the measurement 
for the resistivity in Fig. 4 of \cite{Yusof}
, shows a linear behavior 
in $\omega$ for the $Im(\Sigma)$ and it also shows the 
kink in the dispersion.
It will be very interesting to have experimental determinations of
 the resistivity at lower temperatures,
for such a sample. According to our interpretation the experiment should show 
a linear $T$ dependence of the resistivity due to the MFL 
contribution at low temperature. 

The experimental data for the resistivity as a function of $T$ for 
 optimally doped samples both in $Bi2212$ and 
$YBaCuO$ does not show any curvature\cite{Timusk1}. The behavior is  
surprisingly linear in $T$ with practically no finite 
intercept at zero temperature.
We are forced to  point out that the kink also appears in these samples.
Then, the question is: is $\lambda_{tr}^{ph}$ exactly zero in optimally doped?.

A final discussion on this last point is in order. $1/N$ expansion on 
the t-J model have shown an instability toward a flux phase at a 
critical doping $\delta_c$ (for a given relation $J/t$)\cite{Capellutti}.
In \cite{Capellutti} $\delta_c$ was identified with the optimally 
doped point of the phase diagram of HTC cuprates. In addition, in 
\cite{Kulic} it was shown that the 
 electron-phonon coupling becomes more and more sharp at $\vec{k}=0$
, due to vertex corrections.
This is the main reason for the decreasing of $\lambda_{tr}^{ph}$ when we
approach the flux-phase from the overdoped region\cite{Greco}. If at $\delta_c$
the sharp electron-phonon vertex behaves like a delta function in
$\vec{k}$ , $\lambda_{tr}^{ph}$ will be exactly zero.
 Our previous 
results give a phenomenological support to this speculation.
More theoretical studies will be necessary to clarify this question
\cite{dagotto}.

In summary, we have presented a possible explanation 
of the recent ARPES experiments in cuprates superconductors
by including an EPHI in the context of a MFL
theory. In this way, we have accounted at the same time for the linear 
$\omega$-dependence of the imaginary part of the self-energy and for 
the break in the electron dispersion at a characteristic energy scale 
of 50 meV.
We have confronted our model against the transport data and concluded
 that the EPHI which is relevant for transport is
strongly screened respect to the one extracted from 
ARPES experiments. This condition is supported by strong coupling
theories on the t-J model.

A.G. acknowledges R. Zeyher for useful discussion and Fundaci\'on
 Antorchas for financial support. 
We acknowledge J. Riera for very illuminating discussions
and A. Trumper for critical reading of the manuscript.

\begin{figure}[htb]
\epsfig{file=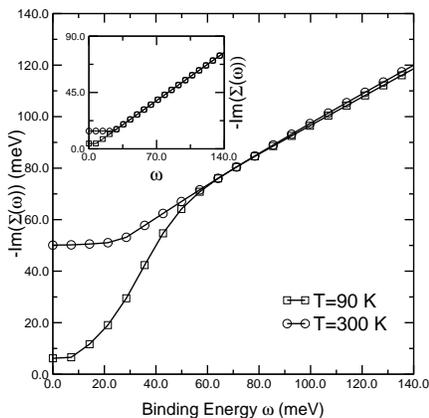,width=6cm}
%\vspace{-2cm}
\caption{
Imaginary part of the electron self-energy. The inset shows the behavior
of a pure MFL system where the two lines collapse in a single one 
at high energy. In contrast, 
in our model the two lines (main figure) split at frequency larger than 
$\sim 80 meV$  due to EPHI.} 
\label{fig1}
\end{figure}

\begin{figure}[htb]
\vspace{-1cm}
\epsfig{file=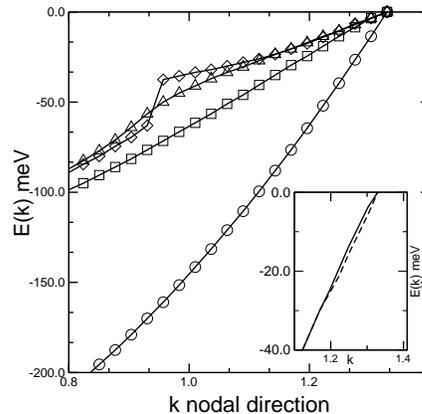,width=6cm}
\caption{
Dispersion relation near the Fermi level in the nodal direction. Circles
are the one-electron  band, the squares give the MFL results as obtained
from the peaks positions of the spectral functions. 
Triangles and diamonds show
the effect of electron-phonon interaction on MFL at 130 and 50 K respectively.
The inset corresponds to the MFL results at T=0 (solid) and T=300K (dotted).} 
\label{fig2}
\end{figure}

\begin{figure}[htb]
%\vspace{-1cm}
%\epsfig{file=fig3.ps,width=6cm,angle=-90}
\epsfig{file=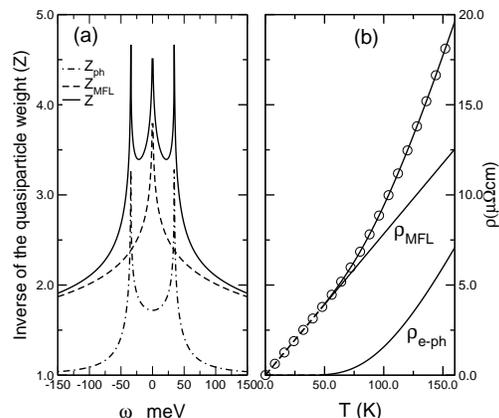,width=6cm}
%\vskip .5 truecm
\caption{
(a) Different contributions to the inverse of the quasiparticle weight. 
(b) Temperature dependence of the electrical resistivity (circles). 
We show separately the contribution of the electron-phonon ($\rho_{e-ph}$) 
 and MFL ($\rho_{MFL}$). 
The dotted line extrapolates the resistivity into the superconducting state.} 
\label{fig3}
\end{figure}

%\end{normalsize}
\end{document}